\documentstyle[preprint,aps,epsf]{revtex}
\tightenlines
\begin{document}
\preprint{TPR-98-34}
\title{On local and global equilibrium in heavy ion collisions}

\author{J. Sollfrank}
\address{Institut f\"ur Theoretische Physik,
D-93040 Universit\"at Regensburg, Germany}

\maketitle

\begin{abstract}
The thermal model is commonly used in two different ways for the 
description of hadron production in ultra-relativistic heavy
ion collision. One is the application of the thermal model to $4\pi$ 
integrated data and the other is the thermal description of central d$N$/d$y$
ratios. While the first method implicitly assumes global
equilibrium the other scenario assumes Bjorken scaling within the 
investigated rapidity range. Both assumptions are only approximations
for real physical collision systems. We study the impact of both
approximations for the extraction of thermal parameters on the exemplary
case of S+S collisions at SPS energies. The particle distributions
are modeled by a hydrodynamical description of the relevant collision 
system.
\end{abstract}

\pacs{25.75.-q,25.75.Dw}

\section{Introduction}

One goal of the studies of high energy collisions is the understanding
of the inclusive hadron production \cite{Satz98}. 
A relevant question is whether the
hadron production is dominated by phase space and statistical
laws or by dynamical constraints of the underlying theory of the
strong interaction. There is strong evidence that even in
elementary collisions like $e^+ e^-$ or pp at sufficiently high energies 
the dominant factor is a statistical filling of phase space 
\cite{Fermi50,Becattini97}.

In heavy ion collisions one may be even closer to thermodynamic
behaviour due to secondary interactions. Therefore a big
effort is going on to study thermal behaviour in these collisions
by microscopic models \cite{Sorge96,Bass98,Bravina98} as well as
to classify directly the experimental final hadronic state 
by a thermal model 
\protect\cite{Davidson,Letessier,Cleymans,Andersen94,BraunMunzinger,Panagiotou,Gorenstein,Spieles97,Sollfrank97a,Sollfrank94,Tiwari96,Becattini98}. 
While within the thermal model applied at low incident energies the 
whole system may be regarded as one 
fireball in approximate global thermal and chemical equilibrium this is not
anymore justified in ultra-relativistic heavy ion collisions.
The incoming nucleons are only partly decelerated and the leading 
protons cannot equilibrate with the mesons in the center.
However, the assumption of local thermal equilibrium might
still be a valid assumption in these reactions.
 
If there is only local thermal equilibrium a fundamental problem arises
in analyzing particle yields directly from experiment. 
A measured hadron does not tell us from which spatial region it comes from.
In principle a model is needed which provides the spatial information
about the particle production, like the hydrodynamic model
or so called event generators.
On the other side one may try to analyze chemical equilibration of 
particle production in ultra-relativistic heavy ion collisions directly 
from the measured particles without a detailed model about spatial
differences in thermodynamic quantities. This is done in two
different ways where both methods make necessarily some compromise.

The first method is to restrict oneself to a limited region in
momentum space (e.g. around midrapidity) and to assume that the considered
region in rapidity corresponds to a homogeneous spatial region 
in thermal equilibrium, i.e.~all thermal parameters are constant over that 
region
\cite{Davidson,Letessier,Cleymans,Andersen94,BraunMunzinger,Panagiotou,Gorenstein,Spieles97,Sollfrank97a}.
This procedure has the disadvantage that there is no one-to-one 
correspondence of spatial regions to regions in momentum 
(rapidity) space. One local cell contributes in general to different 
rapidity regions and even differently for particles of different mass. 
However, in the limit of infinite collision energy we have Bjorken 
scaling \cite{Bjorken83} 
which assures that along longitudinal proper time contours no spatial 
gradients of local variables are present. (There may still be
gradients in transverse direction which we assume to be small). 
We recover global equilibrium
again. Note that we use here and in the following the notion
"global equilibrium" if spatial regions have the same values of all
intensive thermodynamic variables even if there is no causal
connection between these regions. In the scaling  limit
a thermal and chemical analysis of particle spectra directly from
experiment is possible again and in the case of RHIC and LHC this
may be the only reasonable way to do it. 

At SPS energies and below the rapidity spectra don't show Bjorken scaling.
Therefore a chemical analysis in a limited rapidity region might
be questionable \cite{Sollfrank98b}. We will investigate here how
well the Bjorken scaling assumption in connection with a chemical
analysis works using a hydrodynamic model at SPS energies.

The mentioned problem of rapidity cuts leads to the second method for a 
chemical analysis, i.e. a global thermal model to $4\pi$ integrated data
\cite{Sollfrank94,Tiwari96,Becattini98}.
This method implicitly assumes global chemical equilibrium in the above 
mentioned sense.
This assumption is less and less justified the smaller the colliding
nuclei are and the higher the collision energy is. In S+S collisions
at CERN-SPS a clear deviation from global equilibrium between pions
and net protons is seen in that rapidity spectra \cite{Alber98}. 
In our study here we first like to address the question about the error 
of a global thermal
fit to $4\pi$ integrated data in the case of only local equilibrium.

\section{Local and global equilibrium}
Let us assume that particles decouple (freeze-out)
from spatial regions which build up a continuous 3-dim
freeze-out surface $\Sigma^\mu$ embedded in the 4-dim space-time. Along this 
surface we assume further to have local thermal and chemical equilibrium,
but the thermal parameters may vary along this surface. The local particle
densities $\rho_i(x)$ are then given by the local thermal parameters,
temperature $T(x)$, baryon chemical potential $\mu_{\rm B}(x)$,  
and strangeness chemical potential $\mu_{\rm S}(x)$:
\begin{equation} \label{density}
  \rho_i (x) =  (2J_i+1)\, 
\frac{T(x)}{2\pi^2} \sum_{j=1}^{\infty} (\mp 1)^{j+1} \; 
\exp[j(\mu_{\rm B}(x)B_i + \mu_{\rm S}(x)S_i)] \;  
  \frac{m_i^2}{j} \; {\rm K}_2(\frac{j m_i}{T(x)}) \; , 
\end{equation}
where $J_i$ is the spin degeneracy, $B_i$ the baryon number, $S_i$
the strangeness, $m_i$ the mass of particle $i$ and the $-$ ($+$)
sign is for fermions (bosons), respectively.  We neglect
possible suppression factors like strangeness suppression 
$\gamma_{\rm s}(x)$ etc which are needed if only relative chemical
equilibrium is present \cite{Rafelski91}. We also assume that 
strangeness is conserved locally, i.e. the net strangeness density
is zero everywhere. Thus $\mu_{\rm S}(x)$ can be expressed locally
as a function of $T(x)$ and $\mu_{\rm B}(x)$.
The multiplicity of a particle species $i$ is calculated by
\begin{equation}
N_i^{\rm local} = \int\limits_{\Sigma(x)}{\rm d}^3 \sigma_\mu(x) \; 
j_i^\mu[T(x),\mu_{\rm B}(x),\mu_{\rm S}(x)] \; ,
\end{equation}
where $j_i^\mu$ is the particle current of species $i$.
We use the definition of Eckart \cite{Eckart40} for
the four velocity $u^\mu$ and we can therefore decompose the current
$j_i^\mu(x) = \rho_i(x) u^\mu(x)$ into the product of the local
rest frame density  $\rho_i(x)$ and the four velocity $u^\mu(x)$.

In the case of a global thermal fit one assumes that all thermal
parameters are constant and 
thus the multiplicities are given by
\begin{equation} \label{global}
N_i^{\rm global} (T,\mu_{\rm B},\mu_{\rm S})
 = \int\limits_{\Sigma(x)}{\rm d}^3 \sigma_\mu(x) \; 
\rho_i(T,\mu_{\rm B},\mu_{\rm S}) \; u^\mu(x) 
= \rho_i(T,\mu_{\rm B},\mu_{\rm S}) \; V \; ,
\end{equation}
where $V = \int_{\Sigma}{\rm d}^3 \sigma_\mu(x) \; u^\mu(x)$ is
the Lorentz invariant comoving eigen volume.
We define the freeze-out average of a thermal parameter $\cal O$ by 
\begin{equation}
\langle {\cal O} \rangle = \frac{
\int_{\Sigma}{\rm d}^3 \sigma_\mu(x)\; u^\mu(x)\; {\cal O}(x)}{V}\;. 
\end{equation}
Next we define the global fit thermal parameters ${\cal O}^{\rm fit}$
as the parameters which give the minimum $\chi^2$ to experimental data 
or in our study case to the $N_i^{\rm local}$ by
\begin{equation}
\chi^2(V^{\rm fit},T^{\rm fit},\mu_{\rm B}^{\rm fit}) = \sum\limits_i \;
\frac{\left[ N_i^{\rm global} (V^{\rm fit},T^{\rm fit},\mu_{\rm B}^{\rm fit}) 
- N^{\rm local}_i \right]^2}{\sigma_i^2} 
= {\rm Min}\;\chi^2(V,T,\mu_{\rm B}) \;.
\end{equation}

For small variations of the freeze-out parameters along the freeze-out
surface we have the following approximate relation:
\begin{equation}\label{approximation}
N^{\rm local}_i ({\rm d}^3 \sigma_\mu(x)) \approx N^{\rm global}_i
\left(T^{\rm fit},\mu_{\rm B}^{\rm fit},
\mu_{\rm S}^{\rm fit}\right) \approx 
N^{\rm global}_i
\Big(\langle T\rangle,\langle \mu_{\rm B}\rangle,
\langle \mu_{\rm S}\rangle \Big) \; .
\end{equation} 
However, in general all three quantities of Eq.~(\ref{approximation})
are different. 
Here we like to study the validity of Eq.~(\ref{approximation}) for a 
realistic example in order to get a feeling about the goodness of
a global fit to particle yields which arise from a system which is
only in local but not in global equilibrium.

\section{The hydrodynamical study case of S+S collisions}

The minimal deviation from global equilibrium forced by the
experimental spectra is that the baryon density differs locally in space.
So far there is no convincing evidence of a local change in 
temperature and therefore it is usually assumed to be constant
(but see also \cite{Csorgo96}).
A realistic freeze-out surface with constant $T$ but varying baryon 
density and thus varying $\mu_B$ and $\mu_S$
is naturally provided by a hydrodynamical simulation \cite{Rischke98}.
Hydrodynamics by definition assumes local thermal and chemical 
equilibrium. 

We take as an example for our studies S+S collisions at CERN-SPS
since this collision system shows the largest discrepancies in the
proton and pion rapidity distribution and therefore deviations from
global equilibrium. 
We perform a hydrodynamical simulation of S+S collisions with the
same initial conditions and in the same way as described in reference
\cite{Sollfrank98a}. 
The only difference is that the freeze-out hypersurface
is defined now on the contour of constant temperature $T = 140$ MeV. The
temperature and the chemical potentials follow from the local
energy density and baryon density with the help of the used equation
of state which was defined in \cite{Sollfrank97b} and labeled as EOS A. This
equation of state contains very few hadronic resonances. In order
to have a typical resonance spectrum for a chemical analysis 
we use the resonance spectrum up to a mass of 1.7 GeV for
calculating the particle spectra. This introduces a small inconsistency since
the equation of state in the hydrodynamical evolution is different
from the equation of state used for particle spectra. However,
at the low freeze-out temperature of 140 MeV
the higher resonance states are of minor importance. We show
that the calculated freeze-out hypersurface is still compatible
with the higher number of resonance states by comparing the resulting
spectra with experiments. In Figure \ref{spectra} we show as solid lines the
$h^-$ and net proton rapidity and transverse mass distributions.
The spectra are calculated as described in \cite{Sollfrank97b}
using the description of Cooper and Frye \cite{Cooper74}. 
We see that the calculated spectra are still in reasonable agreement 
with the data despite the use of the larger resonance input.

\begin{figure}[t]
\caption[]{Distribution of sub-volumes $V$ having a given freeze-out
baryon chemical potential $\mu_{\rm B}$. \label{mub}}
\hspace{0cm}\epsfxsize 12cm \epsfbox{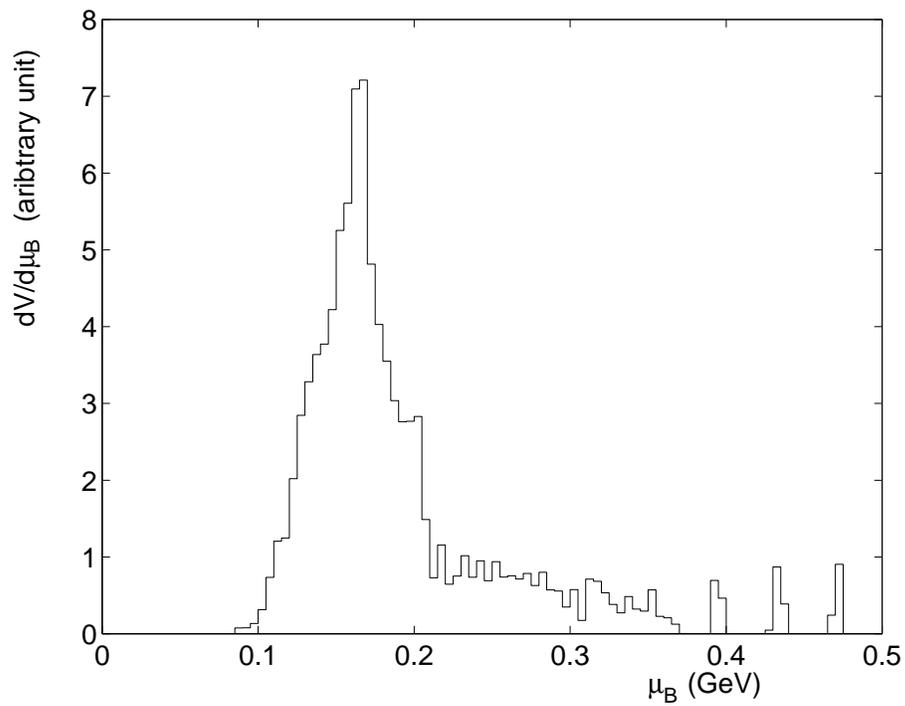}
\end{figure}

We now have a model system which is clearly out of global 
chemical equilibrium. In order to show the deviations from
global equilibrium, we plot in Figure \ref{mub} the distribution
of sub-volumes ${\rm d}V/{\rm d}\mu_{\rm B}$ as function of
$\mu_{\rm B}$ as they result from our hydrodynamical simulation.
The width in $\mu_{\rm B}$ is of order 100 MeV around the
average of $\langle \mu_{\rm B} \rangle = 193$ MeV.
We see a large spread in $\mu_B$ indicating
that there is a large deviation from global chemical equilibrium
with respect to baryon number.

The resulting spectra are integrated over $4\pi$ and the multiplicities 
$N_i^{\rm local}$ are shown in Table \ref{fittable}. In order to perform 
a $\chi^2$ fit we give these yields a relative error as typical for 
experiment \cite{Sollfrank98b} shown in brackets in Table \ref{fittable}.

In a first attempt to describe the local yields one may take
a global equilibrated thermal model with parameters resulting from
averaging over the freeze-out hypersurface. We first take
as chemical potential for the global description the averaged
values $\langle \mu_{\rm B}\rangle$ and
$\langle \mu_{\rm S}\rangle$. The result 
is given in the third column of Table \ref{fittable}.
All yields are underpredicted. Mesons come out right but baryons
and anti-baryons yields are too small. The reason is that
baryon yields are proportional to the baryon fugacity 
$\lambda_{\rm B} = \exp(\mu_{\rm B}/T)$ and
$\langle \mu_{\rm B}\rangle \le T \ln  
\langle \lambda_{\rm B}\rangle$. In the case of the 
anti-baryons we have
$\exp(-\langle \mu_{\rm B} \rangle/T) \le\langle \lambda_{\rm B}^{-1}\rangle$.

Using the average chemical potential leads in a global model to a
reduction in the total baryon number. In order to avoid this problem
one may use $\langle \lambda_{\rm B}\rangle$ and 
$\langle \lambda_{\rm S}\rangle$ as parameters in the global model.
Since the baryon yields are proportional to $\lambda_{\rm B}$ in
Boltzmann approximation the
local and global numbers for non-strange baryons are the same up
to minor corrections due to resonance decays and Fermi statistic.  
The result of such a calculation is shown in the fourth column
of Table \ref{fittable}. Such a scenario, however, leads to large
discrepancies for the anti-baryons. Therefore such a description is not
satisfactory, either.

Now we perform a fit to the local yields with a global thermal model.
We take as fit parameters the volume $V$, temperature $T$, and the 
baryon chemical potential $\mu_{\rm B}$. $\mu_{\rm S}$
is determined by the requirement of strangeness neutrality and
not used as a fit parameter.
The result of the fit is shown in Table \ref{fittable}, too. We recover
in this fit nearly the input temperature and get a 
$\mu_{\rm B}^{\rm fit} = 205.7$ MeV 
which is between the average $\langle \mu_{\rm B} \rangle = 193.2$ and
the $\mu_{\rm B}$ resulting from the average 
$\langle \lambda_{\rm B} \rangle$, $\mu_{\rm B} = 218.4$ MeV. 
The deviations of individual yields of the global fit from the
local integrated ones are small. The average deviation is of order 4\%.
The largest deviations are of order 10\% for the anti-nucleons and 
the $\Omega$.

We conclude that the performance of a global thermal fit to
$4\pi$ integrated data is fine because the deviations in 
temperature and volume from the exact numbers are small
in the studied case of S+S collisions at SPS energies. We
expect that going to larger nuclei and to smaller energies
the amount of stopping increases and therefore the assumption 
of global equilibrium for extracting thermal parameters is 
even more reliable.

\section{Rapidity cuts}

\begin{figure}[t]
\caption[]{Comparison of particle spectra of a
hydrodynamical simulation with freeze-out at constant temperature
$T = 140$ MeV (solid line) with experimental data of
NA35 \cite{Bachler94}. The dotted line results if all fluid cells
at freeze-out is given by hand a constant 
$\mu_{\rm B}(x) = $ 218.4 MeV and corresponds
therefore to a scenario of global equilibrium. 
The dashed line correspond to the spectra of a static fireball
with the same volume as the hydrodynamical simulations as well as
$T = 140$ MeV and $\mu_{\rm B} = $ 218.4 MeV. 
\label{spectra}}
\begin{minipage}{7cm}
\epsfxsize 7cm \epsfbox{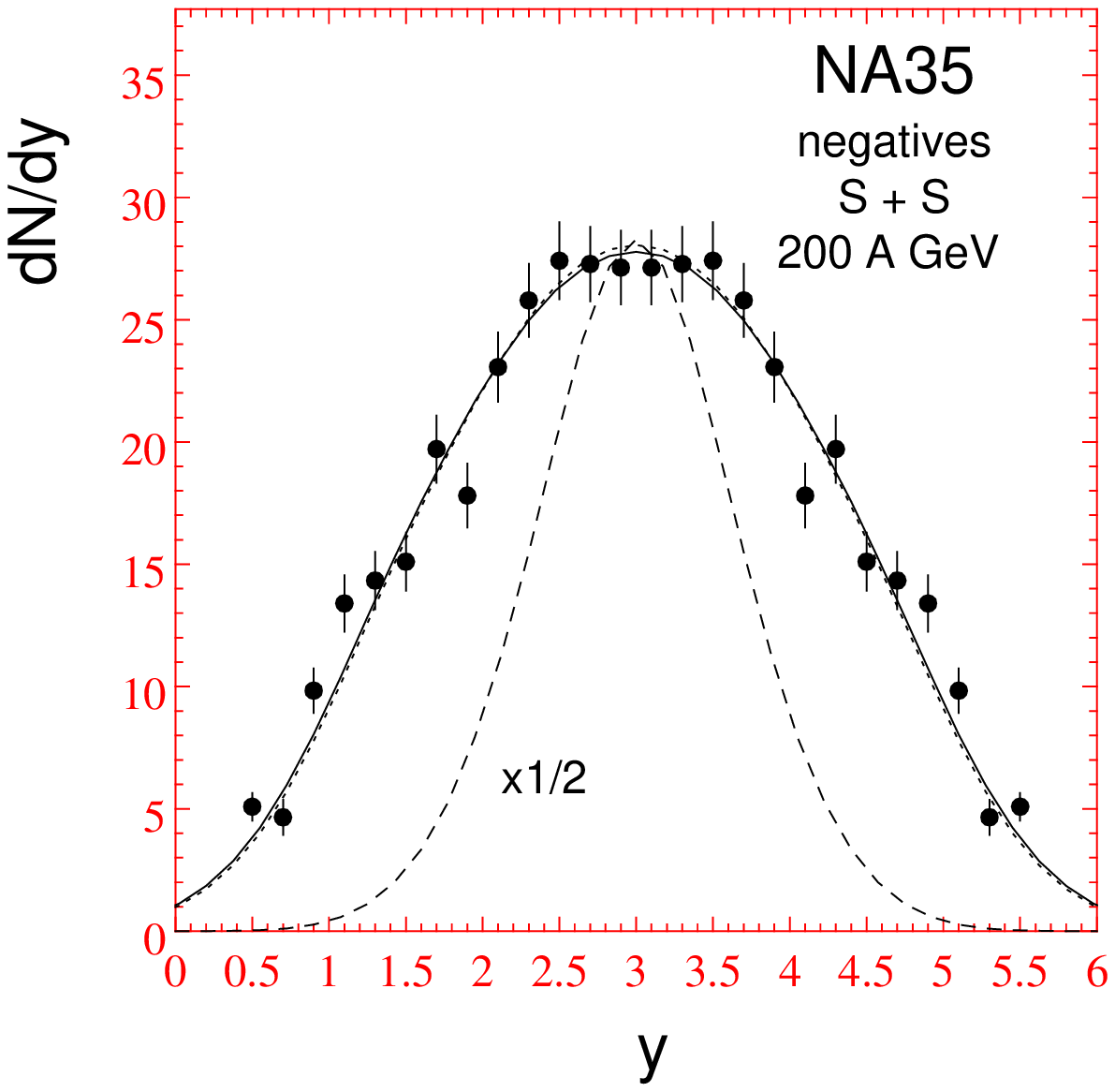}
\end{minipage}
\hfill
\begin{minipage}{7cm}
\epsfxsize 7cm \epsfbox{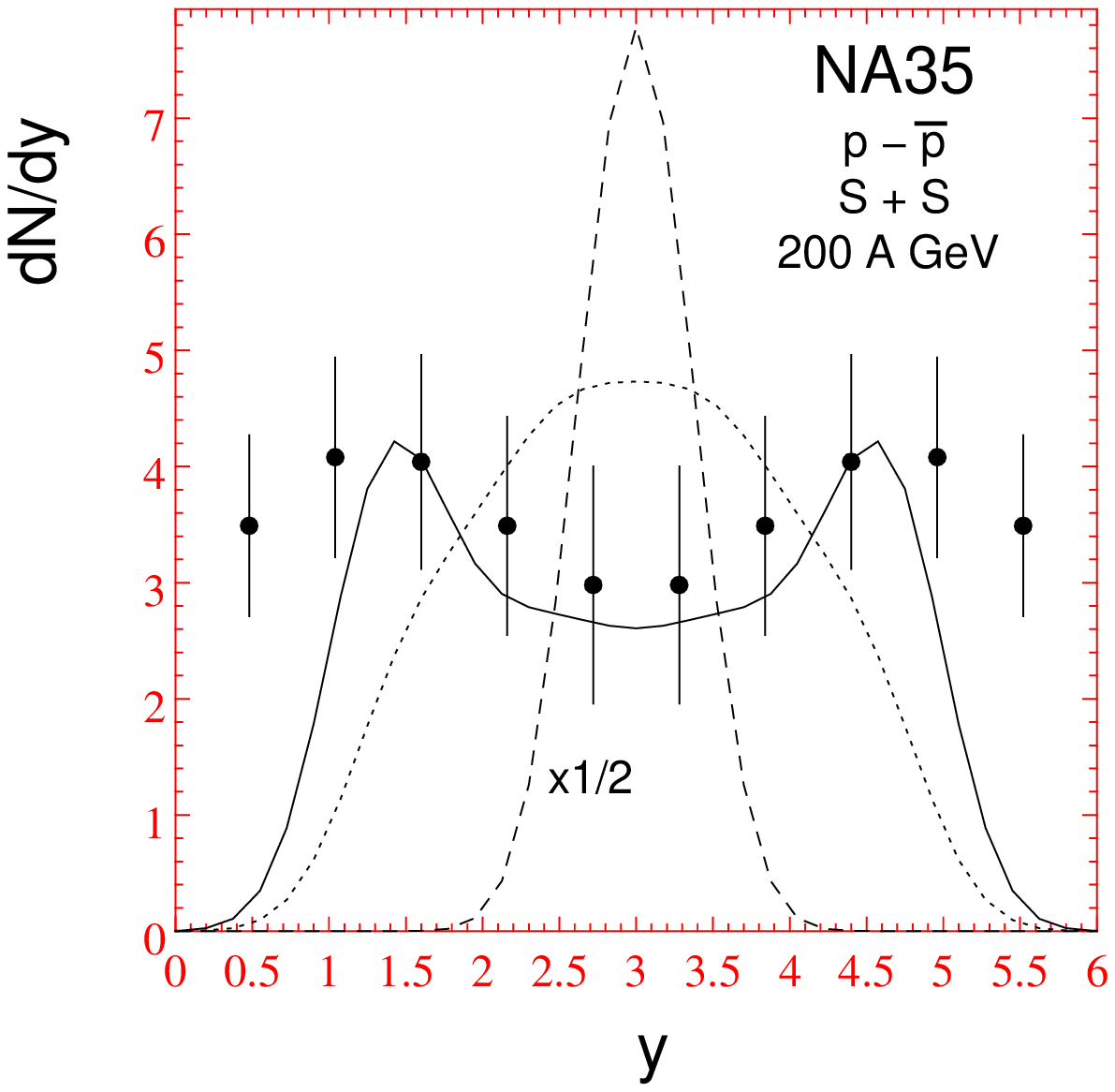}
\end{minipage}

\begin{minipage}{7cm}
\epsfxsize 7cm \epsfbox{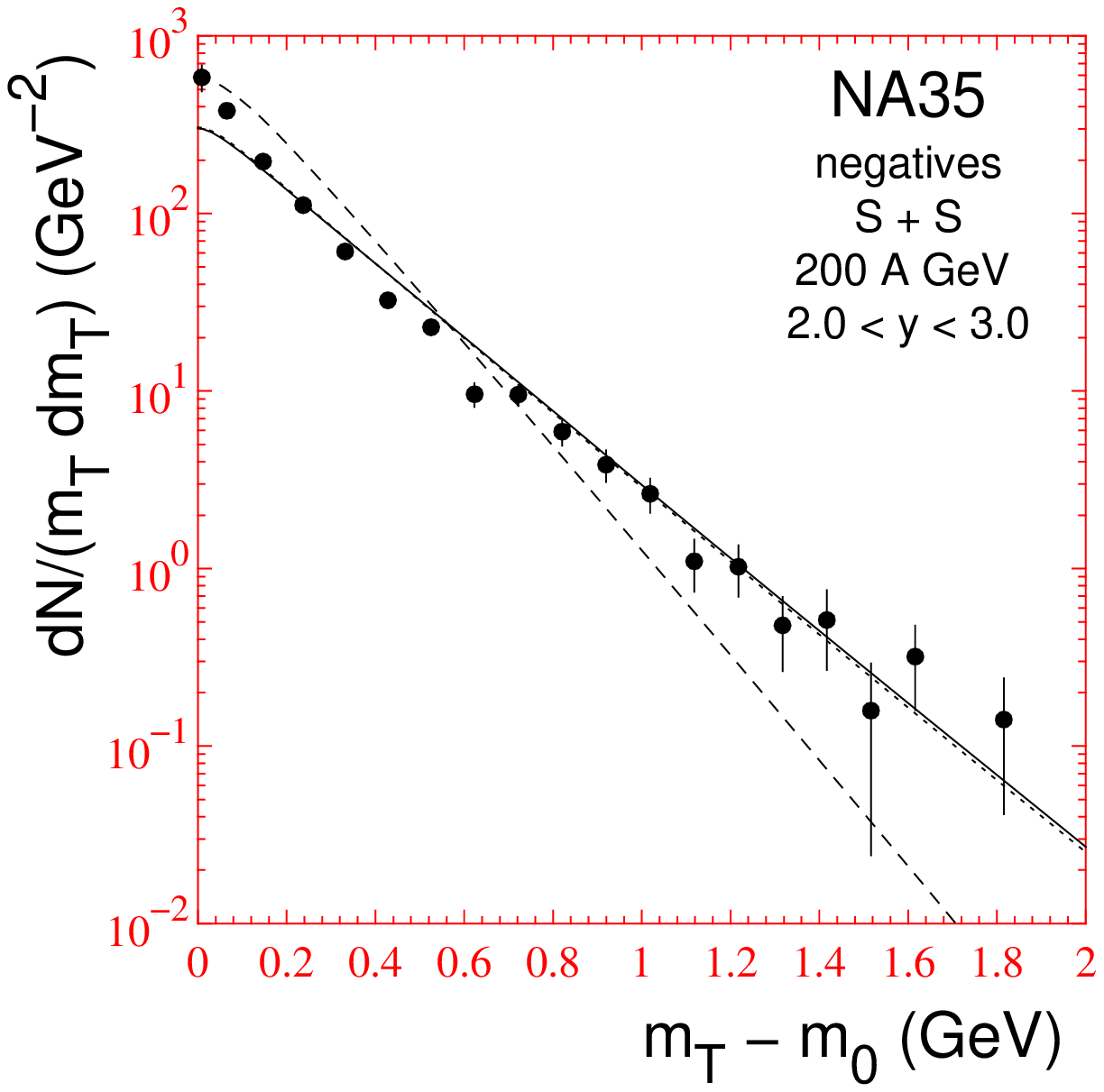}
\end{minipage}
\hfill
\begin{minipage}{7cm}
\epsfxsize 7cm \epsfbox{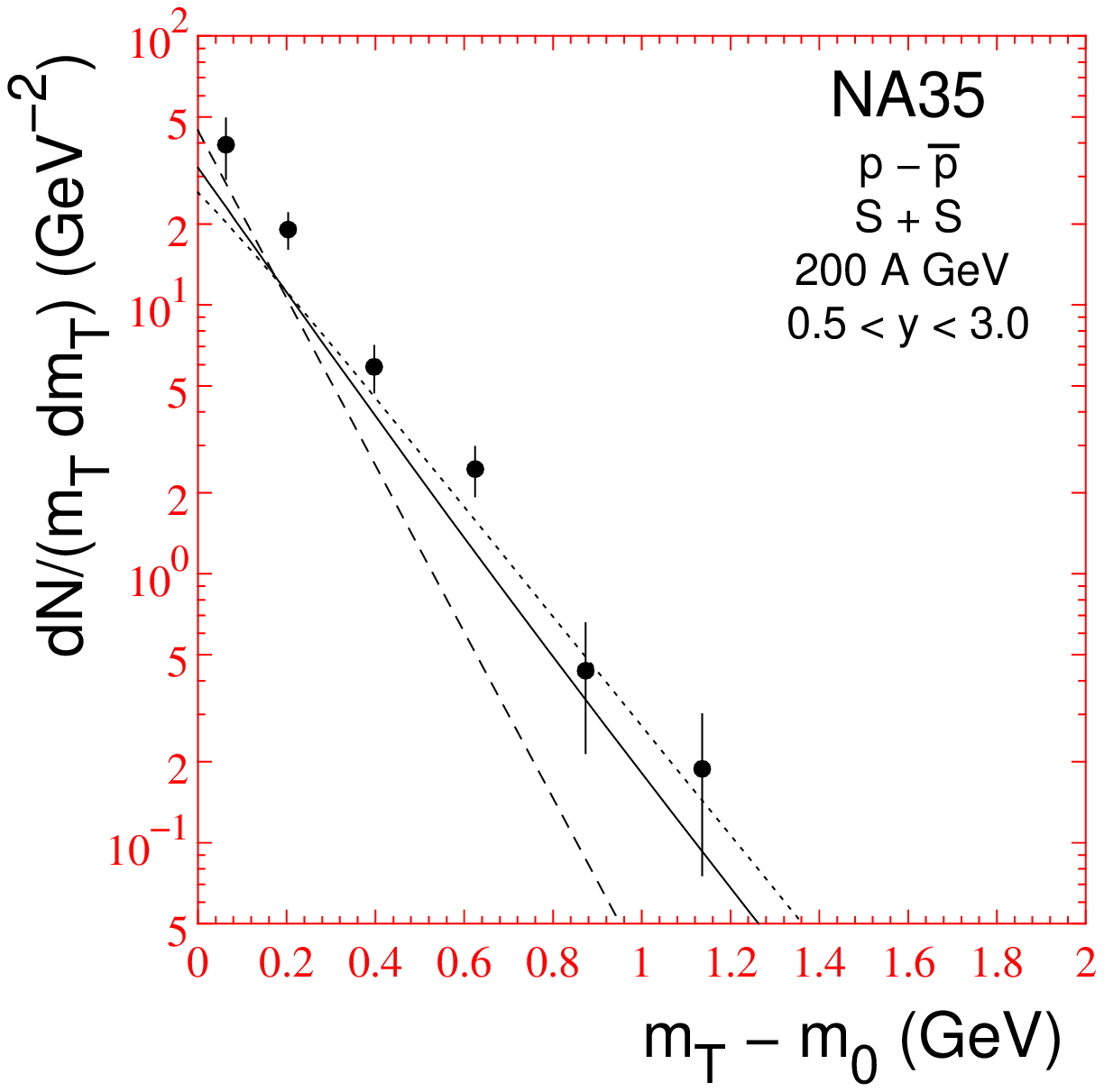}
\end{minipage}
\end{figure}

Next we study the influence of cuts in rapidity
on the extraction of thermal parameters. For all particles in
Table \ref{fittable} we integrate the corresponding 
spectra of the hydrodynamical simulation only over a finite
interval in rapidity  similar to our studies in \cite{Sollfrank98b}.
The resulting particle yields are fitted in the same way as done
before in case of $4\pi$ yields. This means that 
we assume Bjorken scaling in the sense that the multiplicities
of particles in a finite rapidity range are still given by
Eqs.~(\ref{global},\ref{density}).
The resulting thermal parameters are shown in Table \ref{smallone}.
Before discussing the result of that exercise we construct
two hypothetical cases for particle production in order
to compare with.
First we take the result of the hydrodynamical simulation
and give
every fluid cell on the freeze-out surface by hand a constant 
$\mu_{\rm B} = 140 \;{\rm MeV} \times \ln{\langle \lambda_{\rm B}\rangle} 
= 218.4$ MeV. Then we have a system in global equilibrium, but still 
exhibiting the same flow in longitudinal and transverse direction as in
the hydrodynamical simulation. Some of the resulting spectra are shown
in Figure \ref{spectra}. The spectra are integrated over finite
rapidity intervals and fitted with the global thermal model
as done before. The resulting thermal parameters are shown in
Table \ref{smalltwo}. Table \ref{smallthree} shows the result
of the same procedure applied to the spectra of a static fireball
with the same volume as the hydrodynamical simulation. The spectra
are also shown in Figure \ref{spectra}.

In case of the static fireball with no flow, we see the largest
influence of the cuts. The extracted temperature and $\mu_{\rm B}$
change considerably. Also the $\chi^2$ increases drastically going
to smaller rapidity windows. We conclude that a thermal analysis in a
limited rapidity interval for a static fireball is unreasonable
if the analyzed rapidity interval is smaller than the thermal width
of the lightest particle. 

In the case of global equilibrium with longitudinal flow 
the picture changes. Even though we don't see Bjorken scaling in the rapidity 
spectra of Figure \ref{spectra} the extracted thermal parameters are
rather constant and the quality of the fit stays acceptable
even for the smallest rapidity window (see Table \ref{smalltwo}). 
We see a tendency of increasing
temperature and $\mu_{\rm B}$ with decreasing rapidity interval. This
increase is artificially induced by the rapidity cuts but much
weaker than in the case of the static fireball. Since at AGS energies
and especially at SIS energies we expect less longitudinal flow, 
the artificial increase of the fitted temperature
due to rapidity cuts around midrapidity may be larger.

In the hydrodynamic case of Table \ref{smallone} we see a drastic decrease of 
$\mu_{\rm B}$ due to the baryon hole at midrapidity. The temperature,
however, shows a very similar, only very  small increase as in the case of 
global equilibrium with flow and may therefore be attributed to an
artificial increase due to rapidity cuts.  The quality of the fit
is rather independent of the cut. In the same exercise 
\cite{Sollfrank98b} with the
yields from RQMD \cite{Sorge95} we saw a larger increase of
$\chi^2$ with decreasing rapidity window and a larger change of
thermal parameters e.g. temperature. This is due to the fact
that RQMD yields are not in perfect local thermal and chemical 
equilibrium as it
is assumed here. Especially the strange hadron production is
quite different in the central region compared to the fragmentation
regions. In other words,
in the study of rapidity cuts within the RQMD model purely kinematic 
bias on thermal parameters cannot be separated from the impact
of different physics in central regions compared to fragmentation
regions. This is different from our study here, where any changes
in the thermal parameters of tables \ref{smalltwo} and
\ref{smallthree} are artificial changes due to improper 
kinematic cuts.

We summarize that a thermal fit to yields or ratios in a limited
rapidity region gives reasonable results as long as there is
large enough longitudinal flow.

\section{Conclusion}
We have studied the impact of two approximations often used
in the thermal analysis of experimental particle ratios and yields.
We have done this studies for the exemplary case of S+S collisions
at 200 A GeV. First we have shown that in the case of only local
equilibrium a thermal description of the $4\pi$ integrated data 
by a global thermal model leads to deviations. The reason is that
in general the average particle density 
$\langle \rho_i \rangle$ is different from
the density resulting from average parameters, i.e.
\begin{equation}
\langle \rho_i[T(x),\mu_{\rm B}(x), \mu_{\rm S}(x)] \rangle \ne
\rho_i(\langle T\rangle, \langle\mu_{\rm B}\rangle, 
\langle\mu_{\rm S}\rangle) \; . 
\end{equation}
The deviations
depend on how the average is taken and can be as large as 20\%.

Minimal deviations are achieved by performing a $\chi^2$-fit to the
$4\pi$ integrated yields from a hydrodynamic simulation with
local chemical equilibrium.
Such fits reproduce the constant input 
temperature up to a few percent and lead to a $\mu_{\rm B}$ which
is of the order of the average $\langle \mu_{\rm B} \rangle$.
The deviations are generally small (up to 10\%). The quality of a global
thermal model in case of only local equilibrium in other cases
than studied here, e.g. local variations of temperature, larger
variations in the local baryon density, etc. have to be investigated 
individually. However, we think that the result of a reasonable
description of yields from a only local equilibrated system by 
a global model will to a large extend remain valid.

We also studied the influence of rapidity cuts on the extraction of 
thermal model parameters. We explicitly showed that in a system without 
longitudinal flow the rapidity cuts lead to serious problems.
In the case of S+S collisions at SPS, however, the cuts don't spoil 
the extraction of the temperature but lead to smaller central 
$\mu_{\rm B}$ as it is expected from the dip in proton rapidity spectra. 
We conclude that at SPS energies already enough
longitudinal flow is present to justify the Bjorken scaling assumption,
in which case the fitted thermal parameters are independent of rapidity
cuts.

The decision which of both methods should be used for
a chemical analysis depends on the amount of longitudinal flow
in the system. For low energies and large systems a $4\pi$ analysis 
is recommended while for small systems and high energies the analysis
of d$N$/d$y$ around midrapidity should be done. We have shown that
for S+S collision at SPS energies both methods give reasonable
results. For RHIC and LHC an analysis in the central d$N$/d$y$
is recommended while for lower energies,
especially at GSI we strongly recommend to analyze the $4\pi$ integrated 
data.

\acknowledgments
The work was supported by DFG and BMBF.

\newcommand{\IJMPA}[3]{{\it Int.~J.~Mod.~Phys.} {\bf A#1} (#2) #3}
\newcommand{\JPG}[3]{{\it J.~Phys. G} {\bf {#1}} (#2) #3}
\newcommand{\AP}[3]{{\it Ann.~Phys. (NY)} {\bf {#1}} (#2) #3}
\newcommand{\NPA}[3]{{\it Nucl.~Phys.} {\bf A{#1}} (#2) #3}
\newcommand{\NPB}[3]{{\it Nucl.~Phys.} {\bf B{#1}} (#2) #3}
\newcommand{\PLB}[3]{{\it Phys.~Lett.} {\bf {#1}B} (#2) #3}
\newcommand{\PRv}[3]{{\it Phys.~Rev.} {\bf {#1}} (#2) #3}
\newcommand{\PRC}[3]{{\it Phys.~Rev. C} {\bf {#1}} (#2) #3}
\newcommand{\PRD}[3]{{\it Phys.~Rev. D} {\bf {#1}} (#2) #3}
\newcommand{\PRL}[3]{{\it Phys.~Rev.~Lett.} {\bf {#1}} (#2) #3}
\newcommand{\PR}[3]{{\it Phys.~Rep.} {\bf {#1}} (#2) #3}
\newcommand{\ZPC}[3]{{\it Z.~Phys. C} {\bf {#1}} (#2) #3}
\newcommand{\ZPA}[3]{{\it Z.~Phys. A} {\bf {#1}} (#2) #3}
\newcommand{\JCP}[3]{{\it J.~Comp.~Phys.} {\bf {#1}} (#2) #3}
\newcommand{\HIP}[3]{{\it Heavy Ion Phys.} {\bf {#1}} (#2) #3}
\newcommand{\EPJC}[3]{{\it Euro.~Phys.~J.} {\bf C{#1}} (#2) #3}
\newcommand{\PTP}[3]{{\it Progr.~Theor.~Phys.} {\bf {#1}} (#2) #3}
\newcommand{\etal}{{\it et al.}}


\clearpage
\tighten

\begin{table}
\caption{$4\pi$ particle yields: hydrodynamical simulation of
S+S collisions with local varying baryon density (second column); 
global thermal model using the average $\langle \mu_{\rm B}\rangle$ and 
$\langle \mu_{\rm S}\rangle$ (third column) or
$\langle\lambda_{\rm B}\rangle$ and $\langle \lambda_{\rm S}\rangle$
(fourth column) of the hydrodynamical simulation; 
global thermal fit (fifth column) to the yields of the second column. 
The errors in brackets of column 2 are used for the $\chi^2$ fit
and are motivated by typical experimental errors \protect\cite{Sollfrank98b}.
The numbers in square brackets give the percentage of deviation from
mean numbers of the local yields of column 2.\label{fittable}}
\begin{tabular}{ccrcrcrcr}
particle& 
\multicolumn{2}{c}{$N^{\rm local}$} & 
\multicolumn{2}{c}{$N^{\rm global}$} &  
\multicolumn{2}{c}{$N^{\rm global}$} &  
\multicolumn{2}{c}{$N^{\rm global}$} \\
\cline{1-1}
\cline{2-3}\cline{4-5}\cline{6-7}\cline{8-9}
parameter& 
\multicolumn{2}{c}{hydro yields} & 
\multicolumn{2}{c}{
$\langle \mu_{\rm B}\rangle$,$\langle \mu_{\rm S}\rangle$}&
\multicolumn{2}{c}{
$\langle\lambda_{\rm B}\rangle$,$\langle \lambda_{\rm S}\rangle$}&
\multicolumn{2}{c}{fit} \\ \hline
$p$ &14.95 &($\pm$0.91)& 12.49&[-16.4\%] & 14.95&[0.0\%] & 14.17&[-5.2\%] \\ 
$\bar{p}$ &0.879 &($\pm$0.31)& 0.796&[-9.4\%] & 0.665&[-24.3\%] & 
0.786&[-10.6\%] \\ 
$n$ &14.90 &($\pm$0.91)& 12.45&[-16.4\%] & 14.90&[0.0\%] & 14.12&[-5.2\%]\\ 
$\bar{n}$ &0.875 &($\pm$0.30)& 0.793&[-9.4\%] & 0.663&[-24.3\%] & 
0.783&[-10.5\%]\\ 
$\pi^+$ &81.06 &($\pm$2.51)& 79.83&[-1.5\%] & 80.96&[-0.1\%]& 
80.96&[-0.1\%] \\ 
$\pi^-$ &81.06 &($\pm$2.51)& 79.79&[-1.5\%] & 80.91&[-0.1\%]& 80.91&[-0.1\%]\\ 
$\pi^0$ &90.51 &($\pm$2.81)& 89.13&[-1.5\%] & 90.28&[-0.2\%]& 
90.39&[-0.1\%] \\ 
$K^+$ &16.62 &($\pm$0.53)& 16.33&[-1.7\%] & 16.45&[-1.0\%] & 16.77&[0.9\%]\\ 
$K^-$ &11.34 &($\pm$0.66)& 11.18&[-1.4\%] & 11.13&[-1.8\%]& 11.24&[-0.8\%] \\ 
$K^0_{\rm s}$ &13.62 &($\pm$2.21)& 13.39&[-1.6\%] & 13.43&[-1.3\%] & 
13.64&[0.1\%]\\ 
$\phi$ &1.43 &($\pm$0.19)& 1.43&[-0.2\%] & 1.43&[-0.2\%] & 1.49&[4.2\%]\\ 
$\Lambda$ &5.74 &($\pm$0.61)& 5.20&[-9.3\%] & 6.18&[7.6\%]& 5.89&[2.6\%] \\ 
$\bar{\Lambda}$ &0.541 &($\pm$0.099)& 0.507&[-6.2\%] & 0.428&[-21.0\%] & 
0.513&[-5.1\%]\\ 
$\Sigma^+$ &1.63 &($\pm$0.17)& 1.48&[-9.3\%] & 1.75&[7.6\%]& 1.67&[2.5\%] \\ 
$\bar{\Sigma}^+$ &0.154 &($\pm$0.028)& 0.144&[-6.3\%] & 0.121&[-21.0\%]& 
0.146&[-5.2\%]\\ 
$\Sigma^0$ &1.61 &($\pm$0.17)& 1.46&[-9.3\%] & 1.73&[7.6\%] & 1.65&[2.5\%]\\ 
$\bar{\Sigma}^0$ &0.152 &($\pm$0.028)& 0.142&[-6.3\%] & 0.120&[-21.0\%]& 
0.144&[-5.2\%]\\ 
$\Sigma^-$ &1.57 &($\pm$0.17)& 1.42&[-9.3\%] & 1.69&[7.6\%] & 1.61&[2.6\%] \\ 
$\bar{\Sigma}^-$ &0.148 &($\pm$0.027)& 0.139&[-6.3\%] & 0.117&[-21.0\%] & 
0.141&[-5.1\%] \\ 
$\Xi^0$ &0.659 &($\pm$0.066)& 0.630&[-4.4\%] & 0.742&[12.5\%] & 
0.710&[7.7\%] \\ 
$\bar{\Xi}^0$ &0.097 &($\pm$0.014)& 0.094&[-3.5\%] & 0.080&[-18.1\%] & 
0.097&[-0.1\%]\\ 
$\Xi^-$ &0.644 &($\pm$0.064)& 0.615&[-4.4\%] & 0.725&[12.5\%]& 
0.694&[7.8\%]\\ 
$\bar{\Xi}^-$ &0.095 &($\pm$0.014)& 0.092&[-3.5\%] & 0.078&[-18.1\%]& 
0.095&[-0.0\%]\\ 
$\Omega^-$ &0.073 &($\pm$0.017)& 0.072&[-1.6\%] & 0.084&[14.9\%] & 
0.082&[12.3\%]\\ 
$\bar{\Omega}^-$ &0.017 &($\pm$0.006)& 0.016&[-1.5\%] & 0.014&[-15.7\%] & 
0.018&[6.0\%] \\ \hline
$T$ (MeV) & 
\multicolumn{2}{c}{$\langle 140.0\rangle $} & 
\multicolumn{2}{c}{140.0} &
\multicolumn{2}{c}{140.0} &
141.8&($\pm$1.2) \\
$V$ (fm$^3$) & 
\multicolumn{2}{c}{1220} &  
\multicolumn{2}{c}{1220} &  
\multicolumn{2}{c}{1220} &  
1134 &($\pm$71) \\
$\!\!\mu_{\rm B}$(MeV)$\!\!$& 
\multicolumn{2}{c}{$\langle 193.2 \rangle$} &
\multicolumn{2}{c}{193.2}& 
\multicolumn{2}{c}{218.4} &
205.7&($\pm$5.3)\\ \hline
$\!\!\mu_{\rm S}$(MeV)$\!\!$& 
\multicolumn{2}{c}{$\langle 29.9 \rangle$ }&
\multicolumn{2}{c}{29.9}& 
\multicolumn{2}{c}{31.0}& 
\multicolumn{2}{c}{31.8}\\ 
$\lambda_{\rm B}$ & 
\multicolumn{2}{c}{$\langle 4.758 \rangle$} & 
\multicolumn{2}{c}{3.975} &
\multicolumn{2}{c}{4.758} &
\multicolumn{2}{c}{4.26} \\ 
$\lambda_{\rm S}$ & 
\multicolumn{2}{c}{$\langle 1.248 \rangle$} & 
\multicolumn{2}{c}{1.238} &
\multicolumn{2}{c}{1.248} &
\multicolumn{2}{c}{1.255} \\
$\chi^2$ &  & & 
\multicolumn{2}{c}{19.89/22}& 
\multicolumn{2}{c}{15.35/22} & 
\multicolumn{2}{c}{3.97/22}\\ 
\end{tabular} 
\end{table}

\clearpage
\begin{table}
\caption{\label{smallone}
Thermal fit to particle yields in various rapidity
intervals resulting from a hydrodynamical simulation
of S+S collisions at constant freeze-out temperature of
$T = 140$ MeV corresponding to the solid line
in Figure \ref{spectra}. 
Only the resulting thermal parameters are shown.}
\begin{tabular}{cccccc} 
parameter&{$y\leq$0.5}&{$y\leq$1.0}&{$y\leq$1.5}&{$y\leq$2.0}&{$4\pi$}\\ 
\hline
$T$(MeV)            & {143.4$\pm$1.2} & {143.5$\pm$1.2} & {144.0$\pm$1.2} 
                    & {144.0$\pm$1.2} & {141.8$\pm$1.2} \\ 
$V$(fm$^3$)         & {316$\pm$19}    & {594$\pm$36}    & {793$\pm$49} 
                    & {932$\pm$59}    & {1134$\pm$71} \\ 
$\mu_{\rm{B}}$ (MeV)& {154.8$\pm$5.0} & {165.0$\pm$5.1} & {186.2$\pm$5.2} 
                    & {205.3$\pm$5.3} & {205.7$\pm$5.3} \\ 
$\chi^2$            & {3.40/22}       & {2.45/22}       & {2.64/22}
                    & {3.85/22}       & {3.97/22} \\
\end{tabular} 
\end{table}

\vspace{2.0cm}
\begin{table}
\caption{\label{smalltwo}
Thermal fit to particle yields in various rapidity
intervals resulting from a hydrodynamical simulation
of S+S collisions at constant freeze-out temperature of
$T = 140$ MeV and taking artificially a constant $\mu_{\rm B} = 218.4$
MeV at freeze-out corresponding to the dotted line
in Figure \ref{spectra}.
Only the resulting thermal parameters are shown.}
\begin{tabular}{cccccc} 
parameter&{$y\leq$0.5}&{$y\leq$1.0}&{$y\leq$1.5}&{$y\leq$2.0}&{$4\pi$}\\ 
\hline
$T$(MeV)            & {143.4$\pm$1.2} & {143.4$\pm$1.2} & {143.2$\pm$1.2} 
                    & {142.2$\pm$1.2} & {140.0$\pm$1.1} \\ 
$V$(fm$^3$)         & {314$\pm$20}    & {593$\pm$37}    & {819$\pm$51} 
                    & {1006$\pm$61}   & {1224$\pm$73} \\ 
$\mu_{\rm{B}}$ (MeV)& {226.9$\pm$5.2} & {226.8$\pm$5.2} & {226.2$\pm$5.2} 
                    & {223.9$\pm$5.1} & {218.5$\pm$5.0} \\ 
$\chi^2$            & {4.37/22}       & {3.53/22}       & {3.07/22}
                    & {2.09/22}       & 0.050/22\footnotemark[1] \\ 
\end{tabular} 
{\footnotesize \footnotemark[1]
Here $\chi^2$ should be exactly zero. The finite value is due to the
numerical uncertainty resulting from the integration of discretized 
momentum spectra.}
\end{table}

\vspace{2.0cm}
\begin{table}
\caption{\label{smallthree}
Thermal fit to particle yields in various rapidity
intervals resulting from a static fireball with
$T = 140$ MeV, $\mu_{\rm B} = 218.4$ MeV and $V = 1220$ fm$^3$
corresponding to the dashed line in Figure \ref{spectra}. 
Only the resulting thermal parameters are shown.}
\vspace{0.5cm}
\begin{tabular}{cccccc} 
parameter&{$y\leq$0.5}&{$y\leq$1.0}&{$y\leq$1.5}&{$y\leq$2.0}&{$4\pi$}\\ 
\hline
$T$(MeV)            & {155.1$\pm$1.5} & {144.7$\pm$1.2} & {140.8$\pm$1.1} 
                    & {140.0$\pm$1.1} & {139.8$\pm$1.1} \\ 
$V$(fm$^3$)         & {343$\pm$26}    & {860$\pm$54}    & {1154$\pm$69} 
                    & {1225$\pm$73}   & {1240$\pm$74} \\ 
$\mu_{\rm{B}}$ (MeV)& {259.0$\pm$6.1} & {230.7$\pm$5.3} & {220.6$\pm$5.0} 
                    & {218.7$\pm$5.0} & {218.4$\pm$5.0} \\ 
$\chi^2$            & {61.86/22}      & {13.44/22}      & {1.16/22} 
                    & {0.092/22}      & 0.046/22\footnotemark[2] \\ 
\end{tabular} 
{\footnotesize \footnotemark[2]
Here $\chi^2$ should be exactly zero. The finite value is due to the
numerical uncertainty resulting from the integration of discretized 
momentum spectra.}
\end{table}

\clearpage

\clearpage

\end{document}